\documentclass[3p,twocolumn,sort&compress, fleqn]{elsarticle}
\usepackage{amsmath, bm}
\usepackage{amssymb}
\usepackage{amsmath}
\usepackage{braket}
\usepackage{listings}
\usepackage{cprotect}
\usepackage{lmodern}

\journal{Computer Physics Communications}

\def\G1{{G^{(1)}}}

\newcommand{\Pade}{Pad\'{e} }

\usepackage{natbib}
\usepackage{hyperref}
\hypersetup{
        colorlinks=true,
}
\usepackage{listings}

\long\def\beginmypgfpdfnamed#1#2\endmypgfpdfnamed{\includegraphics{#1}}

\graphicspath{{./pyfig/}}

\begin{document}

\begin{frontmatter}

\title{SpM: Sparse modeling tool for analytic continuation of imaginary-time Green's function}

\author[issp]{Kazuyoshi Yoshimi}\ead{k-yoshimi@issp.u-tokyo.ac.jp}
\author[tohokuuniv1]{Junya Otsuki}
\author[issp]{Yuichi Motoyama}
\author[tohokuuniv2]{Masayuki Ohzeki}
\author[saitama]{Hiroshi Shinaoka} 

\address[issp]{Institute for Solid State Physics, University of Tokyo, Chiba 277-8581, Japan}
\address[tohokuuniv1]{Research Institute for Interdisciplinary Science, Okayama University, Okayama 700-8530, Japan}
\address[tohokuuniv2]{Graduate School of Information Sciences, Tohoku University, Sendai 980-8578, Japan}
\address[saitama]{Department of Physics, Saitama University, Saitama 338-8570, Japan}

\begin{abstract}
We present \verb|SpM|, a sparse modeling tool for the analytic continuation of imaginary-time Green's function, licensed under GNU General Public License version 3.
In quantum Monte Carlo simulation, dynamic physical quantities such as single-particle and magnetic excitation spectra can be obtained by applying analytic continuation to imaginary-time data. However, analytic continuation is an ill-conditioned inverse problem and thus sensitive to noise and statistical errors. 
\verb|SpM| provides stable analytic continuation against noise by means of a modern regularization technique, which automatically selects bases that contain relevant information unaffected by noise.
This paper details the use of this program and shows some applications.
\end{abstract}

\begin{keyword}
Sparse modeling,  Analytic continuation, Imaginary-time/Matsubara Green's function
\end{keyword}

\end{frontmatter}
{\bf PROGRAM SUMMARY}

\begin{small}
\noindent
{\em Program Title:} SpM \\
{\em Journal Reference:}                                      \\
{\em Catalogue identifier:}                                   \\
{\em Licensing provisions:} GNU General Public License version 3\\
{\em Programming language:} \verb*#C++#. \\
{\em Computer:} PC\\ 
{\em Operating system:} Any, tested on Linux and Mac OS X\\ 
{\em Keywords:} Sparse modeling,  Analytic continuation, Imaginary-time/Matsubara Green's function\\ 
{\em Classification:} 4.4  \\ 
{\em External routines/libraries:}  BLAS, LAPACK, and CPPLapack libraries. \\
{\em Nature of problem:} The analytic continuation of imaginary-time input data to real-frequency spectra is known 
to be an ill-conditioned inverse problem and very sensitive to noise and the statistic errors.\\
{\em Solution method:} By using a modern regularization technique, analytic continuation is made robust against noise since the basis that is unaffected by the noise is automatically selected. \\
\end{small}

\section{Introduction}


The method of imaginary-time Green's function is a fundamental framework for computations of quantum many-body systems\cite{Mahan}.
This framework, however, has an intrinsic problem in numerical calculations of spectral functions, which requires analytic continuation from imaginary-frequency to real-frequency domains.
Mathematically, the analytic continuation is classified as an ill-conditioned inverse problem with an infinitely large condition number.
This causes serious problems, in particular, in quantum Monte Carlo simulations\cite{gubernatis_kawashima_werner_2016}: Statistical errors in the imaginary-time quantities are enormously amplified to result in unphysical spectra.
To overcome this problem, several methods have been proposed, including \Pade approximation \cite{PhysRevB.61.5147,PhysRevB.93.075104}, the maximum entropy method \cite{MaxEnt1990, JARRELL1996133,PhysRevB.81.155107},  stochastic methods \cite{PhysRevB.57.10287,PhysRevE.81.056701,PhysRevE.94.063308}, and the neural network method\cite{Yoon2018}. However, a definitive method has yet to be established. 

Some of the present authors have recently proposed a method for analytic continuation that uses the sparse modeling method \cite{PhysRevE.95.061302}. In this method, stable analytic continuation against noise is achieved by selecting bases that are insensitive to the noise. \verb|SpM| is a software package that implements the above method \cite{SpM}. In \verb|SpM|, L1-norm regularization is used to separate relevant information in imaginary-time data from irrelevant information that makes the spectrum unphysical. Furthermore, the solution can be obtained under some constraints, such as the sum rule and non-negativity. 

The rest of this paper is organized as follows. The basic usage of \verb|SpM|, from installation to execution, is described  in Sec. \ref{Sec:BasicUsage}. The algorithm implementation is presented  in Sec. \ref{subsec:formulation}. 
Numerical results and an analysis of output data for a sample file are also demonstrated in Sec  \ref{Sec:Examples}.
The conclusions are given in Sec.\ref{Sec:Summary}.

\section{Basic usage} \label{Sec:BasicUsage}
\begin{figure}[tb!]
  \begin{center}
    \includegraphics[width=1 \columnwidth]{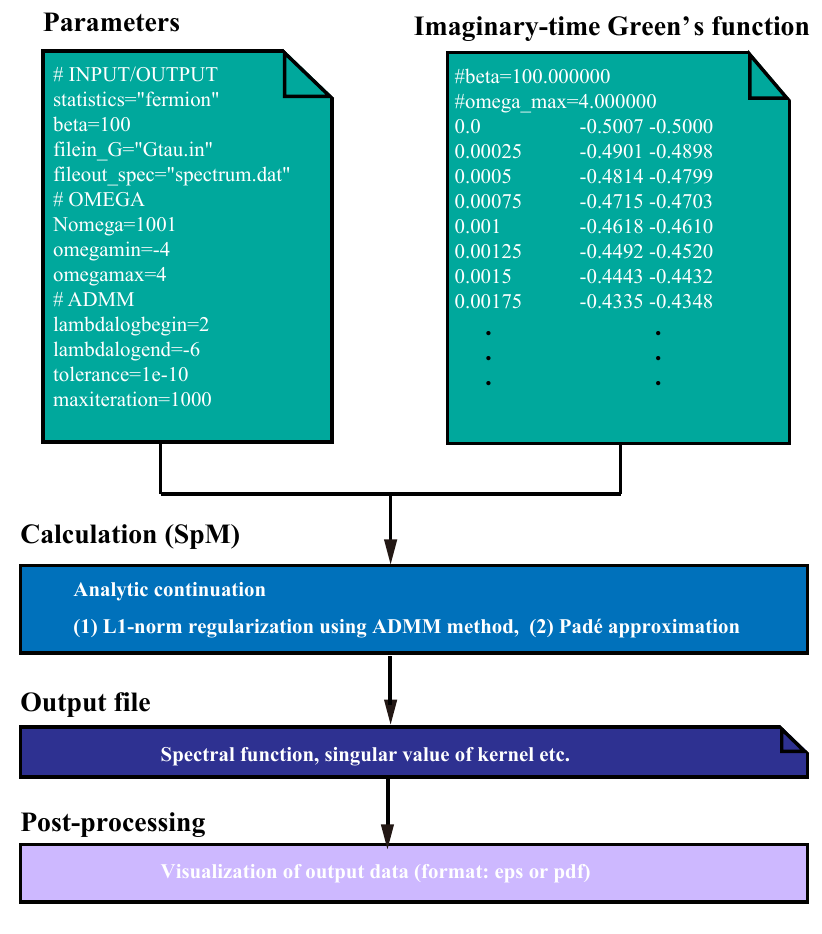}
    \cprotect\caption{Overview of basic usage flow of \verb|SpM|. }
    \label{fig:flow}
  \end{center}
\end{figure}

\subsection{Install}
The stable version of \verb|SpM| can be downloaded from its Github page \cite{SpM}. The gzipped tar file contains the source code and samples. To build \verb|SpM|, a \verb|C++| compiler and the \verb|BLAS|, \verb|LAPACK| and \verb|CPPLapack| libraries are required. The \verb|CMake| utility can be used to manage the build process. An example of the procedure for compiling \verb|SpM| is shown below: 
\begin{verbatim}
$ tar xvzf  SpM-x.y.z.tar.gz
$ mkdir spm.build && cd spm.build
$ cmake spm.src
$ make
\end{verbatim}
Here, \verb|x.y.z| indicates the version number of \verb|SpM| and \verb|spm.src| is a directory that contains the source code.
In this case, the executable file \verb|SpM| is generated in the \verb|spm.build/src| directory.

Some scripts are included for test calculations.
For example, the fermionic spectrum can be obtained by running a test calculation as follows:
\begin{verbatim}
$ cd spm.src/samples/fermion
$ ln -s spm.build/src/SpM.out ..
$ sh ./run.sh
\end{verbatim}
The results can be found in the \verb|output| directory.

\subsection{Usage}\label{subsec:usage}
In this subsection, we explain the usage of \verb|SpM|. 
An overview of the basic usage flow of \verb|SpM| is shown in Fig. \ref{fig:flow}.

\vspace{1em}\noindent
{1. \bf Make a data file}

In \verb|SpM|, the imaginary-time $\tau$ is automatically determined from the inverse temperature \verb|beta|, which is specified in the parameter file, and the total number of steps, e.g. $\tau_i = \frac{\beta}{N_{\tau}}i $. The number of the column where the values of $G(\tau)$ are stored is indicated in the parameter file. An example of the data file is shown in the upper-right corner of Fig. \ref{fig:flow}.

\vspace{1em}\noindent
{2. \bf Make a parameter file}

Parameters defined in this input file 
are classified into the following four sections:\\
(1) {\bf INPUT or OUTPUT}: 
Information regarding input and output data, such as statistics, inverse temperature, and input (output) file name for $G(\tau)$ (spectrum), is specified.
\\
(2) {\bf OMEGA}:
The minimum and maximum values of real frequencies and the total number of target frequencies are given.
\\
(3) {\bf SVD}:
The truncation value of singular values (SV) is given.
\\
(4) {\bf ADMM}:
The parameters for L1 regression, conducted using an algorithm named the alternating direction method of multipliers (ADMM) \cite{Boyd:2011:DOS:2185815.2185816}, are given.

An example of the parameter file is shown in the upper-left corner of Fig. \ref{fig:flow}. For details regarding the input files, please see the official website\cite{SpM}.

\vspace{1em}\noindent
{3. \bf Calculation}

After the two input files have been prepared, \verb|SpM| can be run by typing the following command in the terminal:
\begin{verbatim}
$ spm.build/src/SpM.out -i param.in
\end{verbatim}
Here, \verb|param.in| is the parameter input file.

\vspace{1em}\noindent
{4. \bf Output} 

After calculation, the \verb|output| directory is automatically created.
The following files are output in this directory:\\
(1) \verb|spectrum.dat|: 
Real frequency spectrum $\rho(\omega)$ for the optimal value of the hyper parameter $\lambda$. 
\\
(2)  \verb|lambda_dep.dat|:
Several quantities such as square error in the SV/original basis and L1 norm in the SV basis as a function of $\lambda$.
\\
(3) \verb|find_lambda_opt.dat| :
Auxiliary data that are used to determine the optimal value of $\lambda$. 
\\
(4)  \verb|SV.dat|:
SV of the kernel $K(\tau, \omega)$.\\
The \verb|lambda_opt| directory contains the dataset at the optimized $\lambda$.
The \verb|lambda| directory contains subdirectories \verb|lambda_*| for each $\lambda$, which are used to store the detailed numerical results.

\vspace{1em}\noindent
{5. \bf Post-processing}

To visualize the results, \verb|gnuplot| script files are included in the samples directory.
These script files can be used by typing the following command in the \verb|output| directory:
\begin{verbatim}
$ gnuplot spm.src/samples/plt/*
\end{verbatim}

This command generates four Encapsulated PostScript (EPS) files, namely \verb|spectrum.eps|, \verb|lambda_dep.eps|, \verb|find_lambda_opt.eps| and \verb|SV_log.eps|. 
These scripts can be used by typing the following command in the \verb|lambda_*| directory:
\begin{verbatim}
$ gnuplot spm.src/samples/plt/
lambda_fix/*
\end{verbatim}
For details regarding the output files, please see the official website\cite{SpM}.

\section{Algorithm}\label{subsec:formulation}
\begin{figure}[tb!]
  \begin{center}
    \includegraphics[width=1 \columnwidth]{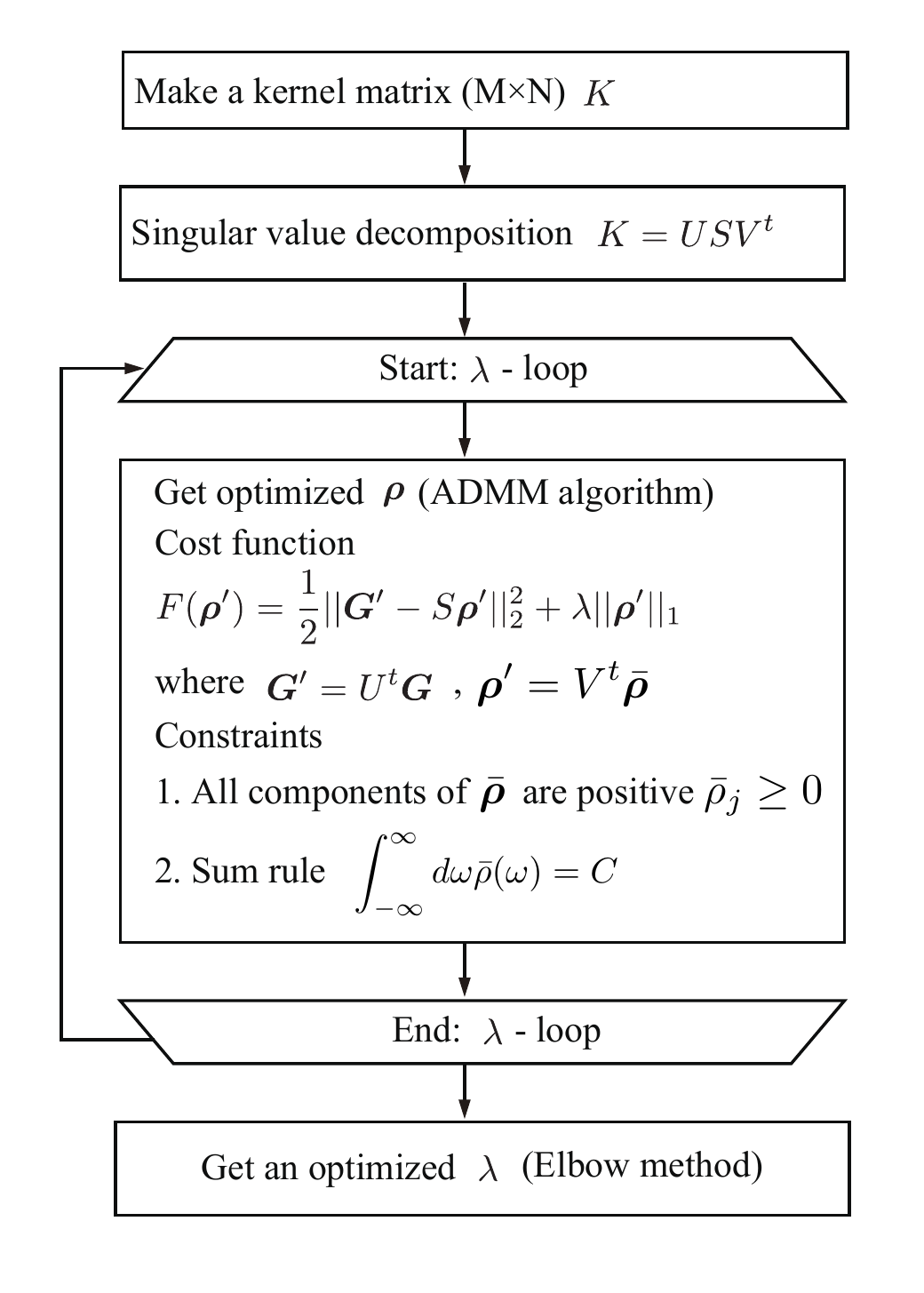}
    \caption{Calculation flow used to obtain spectrum at the optimized $\lambda$. }
    \label{fig:flowSpM}
  \end{center}
\end{figure}

In this section, we explain the calculation flow of \verb|SpM| used to obtain the spectrum using $L_1$-norm regularization.
The calculation procedure consists of the four steps shown in Fig. \ref{fig:flowSpM}. Each step is detailed below.

\vspace{1em}\noindent
{1. Make a kernel matrix}

The input of the analytic continuation is the imaginary-time Green's function $G(\tau)$ computed in the range $0\leq\tau\leq\beta$.
For numerical data,
one may use the exact integral equation between $\bar{\rho}(\omega)$ and $G(\tau)$:~\footnote{Note that the sign of $G(\tau)$ is opposite to that in the ordinary definition. Here, $G(\tau)$ is positive definite in $0\leq\tau\leq\beta$.}
\begin{align}
G(\tau) = \int_{-\infty}^{\infty} d\omega K (\tau, \omega) \bar{\rho}(\omega).
\label{eq:G_K_rho}
\end{align}
Here, $K(\tau, \omega)$ and $\bar{\rho}(\omega)$ are defined as
\begin{align}
K(\tau, \omega) &=
\begin{cases}
\displaystyle \frac{e^{-\tau\omega}}{1 + e^{-\beta \omega}} & (\text{fermion})\\
\displaystyle \frac{\omega e^{-\tau\omega}}{1 - e^{-\beta \omega}} & (\text{boson})
\end{cases},
\label{eq:kernel}\\
\bar{\rho}(\omega) &=
\begin{cases}
\rho(\omega) & (\text{fermion})\\
\rho(\omega)/\omega & (\text{boson})
\end{cases},
\end{align}
where $\rho(\omega) \equiv {\rm Im} G(\omega + i0^+)/\pi$ is the spectrum.

For convenience, we recast Eq.~(\ref{eq:G_K_rho}) into a conventional linear equation as
\begin{align}
\label{eq:y_Kx}
\bm{G} = K \bar{\bm{\rho}}.
\end{align}
Here, the vector $\bm{G}$ is defined as $G_i \equiv G(\tau_i)$, where $\tau_i$ is the $M$-division of $[0:\beta]$. 
The quantities on the right-hand side are defined as $K_{ij} \equiv K(\tau_i, \omega_j)$ and $\bar{\rho}_j \equiv \bar{\rho}(\omega_j)\Delta\omega$, which are obtained after 
replacing the integral over $\omega$ with $N$-point finite differences in the range $[-\omega_{\rm max}:\omega_{\rm max}]$.

\vspace{1em}\noindent
{2. Singular value decomposition}

The singular value decomposition (SVD) of the matrix $K$ is conducted as
\begin{align}
\label{eq:SVD}
K = U S V^{\rm t},
\end{align}
where $S$ is an $M\times N$ diagonal matrix, and $U$ and $V$ are orthogonal matrices of sizes $M\times M$ and $N\times N$, respectively.

\vspace{1em}\noindent
{3. Get the spectrum at fixed $\lambda$}

Introducing new vectors
\begin{align}
\label{eq:SV_basis}
\bm{\rho}'\equiv V^{\rm t}\bar{\bm{\rho}},
\quad
\bm{G}'\equiv U^{\rm t}\bm{G},
\end{align}
we consider cost function $F(\bm{\rho}')$ that includes the $L_1$ regularization term
\begin{align}
\label{eq:F}
F(\bm{\rho}') \equiv \frac12 \| \bm{G}' - S \bm{\rho}'\|_2^2 + \lambda \| \bm{\rho}' \|_1,
\end{align}
where $\lambda$ is a positive constant
and $\| \cdot \|_1$ denotes the $L_1$ norm defined as
$\| \bm{\rho}' \|_1 \equiv \sum_l |\rho_l'|.$
This form of optimization problem is referred to as least absolute shrinkage and selection operators (LASSO)~\cite{10.2307/2346178}.
We solve this optimization problem using the ADMM algorithm developed by Boyd {\it et al.}~\cite{Boyd:2011:DOS:2185815.2185816}.

It is noted that the solution must satisfy two conditions: non-negativity $\bar{\rho}(\omega) \geq 0$ and the sum rule $\int_{-\infty}^{\infty} \bar{\rho}(\omega) d\omega =C$, where $C=1$ for the fermionic case and $ C = \int_{0}^{\beta} d \tau G(\tau) $ for the bosonic case.
These constraints are expressed in terms of the vector $\bar{\bm{\rho}}$ as
\begin{align}
\label{eq:constraint_x}
\bar{\rho}_j \geq 0,\quad \sum_j \bar{\rho}_j = C.
\end{align}
Each constraint can be switched on or off to meet the user's needs.
This is a technical advantage over the maximum entropy method, 
in which the entropy term requires positiveness, $\rho(\omega)>0$~\footnote{For non-negative spectra in the maximum entropy method, see Refs.~\cite{PhysRevB.92.060509} and \cite{PhysRevB.98.205102}.}.
The details of the ADMM method with these constraints are presented in a previous paper \cite{PhysRevE.95.061302}.

\vspace{1em}\noindent
{4. Get the optimal value of $\lambda$}

To find the optimal value of $\lambda$,
we first define the function $f(\lambda)=a\lambda^b$,
which connects the left and right endpoints of $\chi^2(\bm{\rho}';\lambda) \equiv  \frac12 \| \bm{G}' - S \bm{\rho}'\|_2^2 $.
Then, the peak in the ratio $f(\lambda)/\chi^2(\bm{\rho}';\lambda)$ corresponds to the position of the kink in $\chi^2(\bm{\rho}';\lambda)$. 
In this way, we obtain $\lambda_{\rm opt}$.
A similar method was adopted in the literature~\cite{PhysRevLett.112.070603, doi:10.7566/JPSJ.84.024801}.

\section{Examples}\label{Sec:Examples}
\subsection{Calculation of spectrum}\label{ExSpec}

\begin{figure}[tb!]
  \begin{center}
    \includegraphics[width=1 \columnwidth]{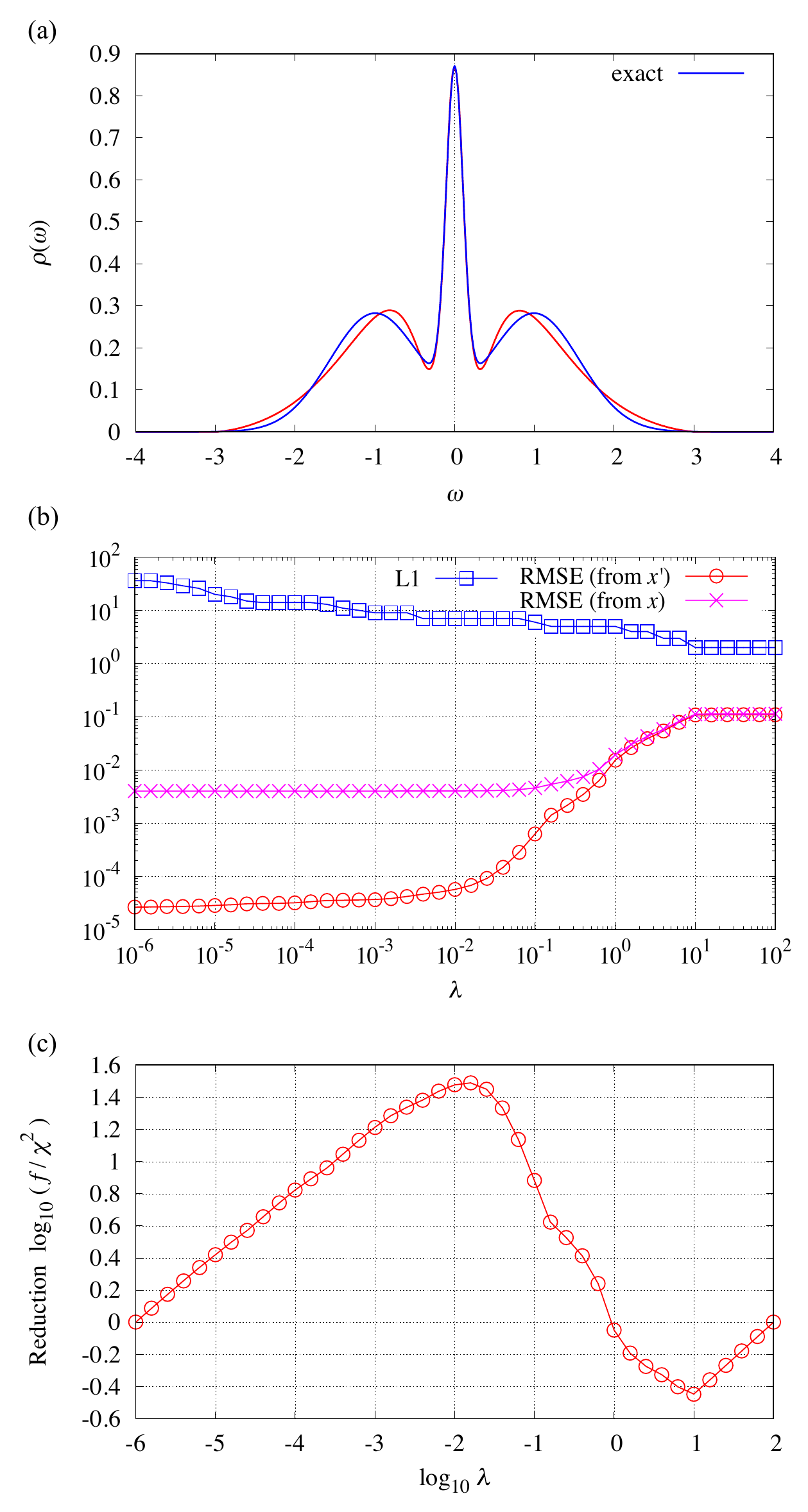}
    \cprotect\caption{Results of the calculation for the sample file in the \verb|fermion/sample| directory. (a) The red line shows the spectrum function obtained by \verb|SpM|. The blue line shows the exact spectrum function. (b) The blue line shows the L1 norm, i.e., the last term in Eq. (\ref{eq:F}). The pink and red lines show the root-mean-squared error (RMSE) obtained in the SV basis ($x'$) and the original space ($x$), respectively. (c) The red line shows the reduction of $\log_{10} (f/\chi^2)$, where $f(\lambda)=a \lambda^b$ and $\chi^2({\bm \rho'};\lambda)\equiv \frac{1}{2}||{\bm G' - S{\bm \rho'}}||_2^2$. Here, $a$ is the slope of the line that connects the left and right endpoints of $\chi^2$.}
    \label{fig:sample}
  \end{center}
\end{figure}

In this section, we first demonstrate \verb|SpM| by calculating the spectrum. The input files are located in the samples directory \verb|spm.src/samples/fermion|.
In this sample, the exact spectrum function, which has three peaks, at $\omega =0, \pm 1$, respectively, is prepared by superimposing three Gaussian functions, as shown in Fig. \ref{fig:sample}(a) (blue line). By using Eq. (\ref{eq:G_K_rho}), the exact imaginary-time Green's function $G_{\rm ex}(\tau)$ is obtained. Then, white noise with standard deviation $\sigma = 10^{-3}$ is added to $G_{\rm ex}(\tau)$. The obtained $G(\tau)$ is output to the file \verb|Gtau.in|.

Next, we prepare the input file to set the parameters as explained in Sec. \ref{subsec:usage}. Here, the sample file \verb|param.in| is as follows:
\begin{verbatim}
# INPUT/OUTPUT
statistics="fermion"
beta=100
filein_G="Gtau.in"
column=1
fileout_spec="spectrum.dat"
# OMEGA
Nomega=1001
omegamin=-4
omegamax=4
# ADMM
lambdalogbegin=2
lambdalogend=-6
tolerance=1e-10
maxiteration=1000
\end{verbatim}
After the above two files have been prepared, \verb|SpM| is executed by typing the following command:
\begin{verbatim}
$ spm.build/src/SpM.out -i param.in
\end{verbatim}

Figure \ref{fig:sample}(a) shows the real-frequency spectrum function (red line). It can be seen that the spectrum function becomes smooth and thus the effect of noise is suppressed by this method. In Fig. \ref{fig:sample}(b), the root-mean-squared error (RMSE) $\chi^2$ in the SV basis is shown by the red line, and that in the original basis is shown by the pink line. The L1 norm corresponding the second term in Eq. (\ref{eq:F}) is plotted as the blue line. The L1 norm monotonically increases with decreasing $\lambda$; i.e., the number of bases that have non-zero coefficients becomes large. The RMSE first decreases toward $\lambda \sim 10^{-1}$, but then becomes saturated in the sufficiently small $\lambda$ region due to over-fitting. 

This over-fitting can be avoided by choosing a suitable $\lambda$. In \verb|SpM|, the elbow method is adopted to get the optimal value of $\lambda$, as explained in Sec. \ref{subsec:formulation}. In this method, the optimal value is determined from the peak position of $f(\lambda)/\chi^2({\bm \rho'};\lambda)$. As can be seen in Fig. \ref{fig:sample}(c), $\lambda_{\rm opt}$ is obtained as $1.585\times 10^{-2}$ for this sample. The spectrum function shown in Fig. \ref{fig:sample}(a) is that at $\lambda_{\rm opt}$.

\subsection{Robustness against noise}
\begin{figure}[tb!]
  \begin{center}
    \includegraphics[width=8cm]{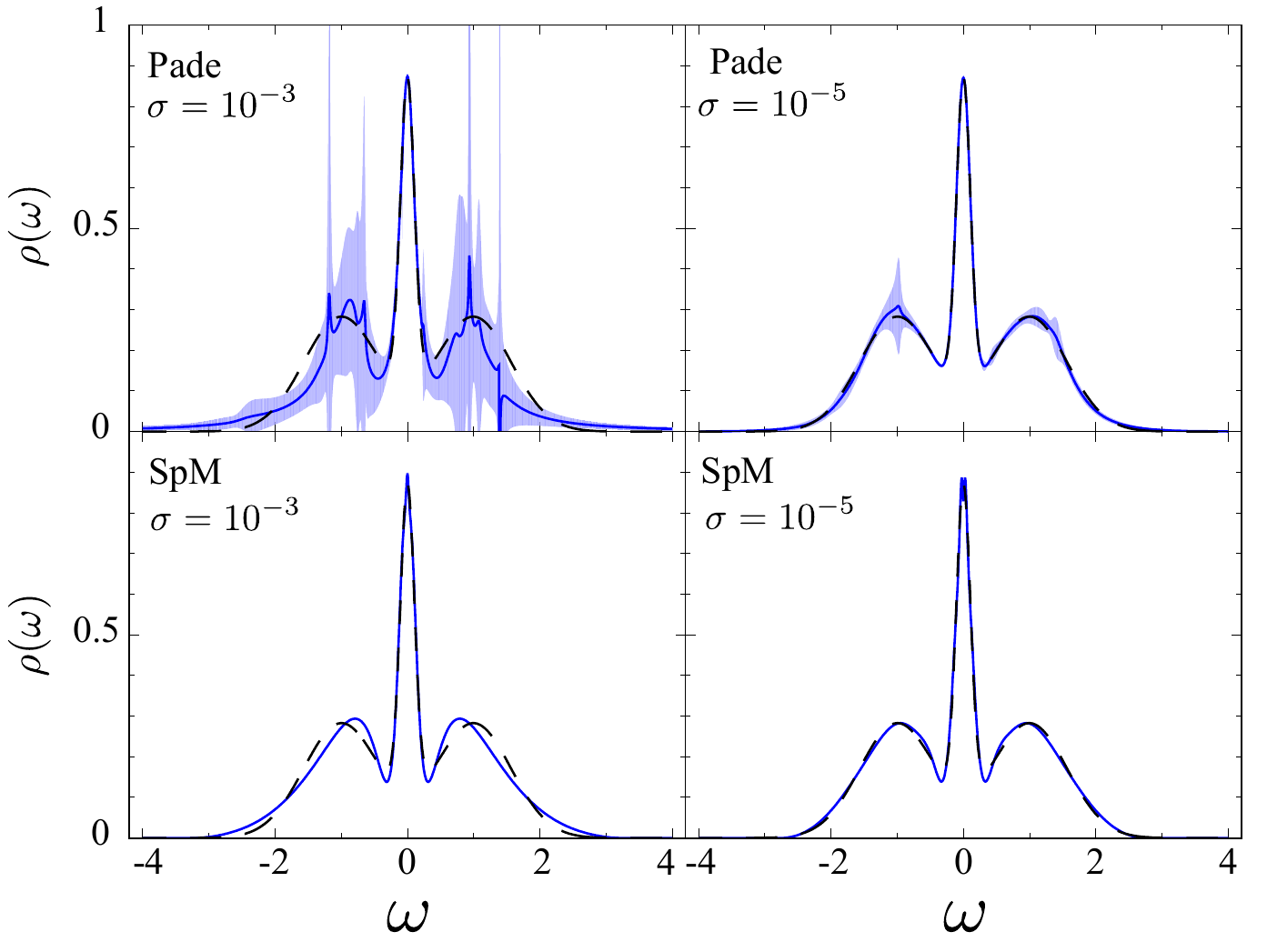}
    \cprotect\caption{Robustness of \verb|SpM| (bottom panels) and \Pade approximation (top panels).
      The black dashed lines represent the ``exact'' spectrum.
      The blue lines and shaded regions denote the mean values and standard deviations, respectively,
      of the estimated spectra from 30 independent Green's functions with white noise $\eta \leftarrow N(0, \sigma)$.
      The left (right) panels show the results under the strong (weak) noise $\sigma = 10^{-3}$ ($\sigma = 10^{-5}$).
    }
    \label{fig:spm_pade_means}
  \end{center}
\end{figure}

Finally, we demonstrate the robustness of \verb|SpM| against the noise of the Green function. For comparison, we also show the results of \Pade approximation.
\Pade approximation is one of the standard methods for analytic continuation \cite{PhysRevB.61.5147,PhysRevB.93.075104}.
In this method, the Green's function in Matsubara frequency $G(i\omega_n)$ is approximated by a rational function $\tilde{G}(i\omega)$ and the spectrum is calculated using analytic continuation as $\rho(\omega) = -\mathrm{Im} \tilde{G}(\omega+i\delta)/\pi$.
Since $\tilde{G}(i\omega)$ is obtained along the imaginary axis, the accuracy of the analytic continuation is high around the zero frequency. Accuracy decreases with distance away from the zero frequency. Thus, it is expected that the effect of noise will be strong in high-frequency regions.

To evaluate robustness against noise, we first prepared the ``exact'' fermionic spectrum depicted in Fig.~\ref{fig:sample}(a) and generated 30 independent Green's functions with white noise $\eta \leftarrow N(0, \sigma)$ as samples.
Next, we reconstructed the spectrum from these samples using \verb|SpM| and \Pade approximation and estimated the mean and standard deviation of the spectrum at each frequency.
Figure~\ref{fig:spm_pade_means} shows the numerical results.
The black dashed curve depicts the ``exact'' spectrum, and
the blue lines and the shaded regions represent the mean values and standard deviations, respectively.
At $\sigma = 10^{-3}$, the effect of noise is large around the peak frequencies at $\omega=\pm 1$ for \Pade approximation. For \verb|SpM|, although the spectrum is stable against noise, the peak frequencies shift from $\omega=\pm 1$. 
At $\sigma = 10^{-5}$, the effect of noise becomes small but the deviation around $\omega=\pm 1$ remains for \Pade approximation. For \verb|SpM|, the shape of the spectrum matches that of the exact spectrum. 
These results indicate that \Pade approximation is a good approximation around $\omega= 0$ but becomes sensitive to noise in the region far from $\omega= 0$, as expected. In contrast, \verb|SpM| is robust against noise. However, the noise moves the peak frequencies relative to the exact ones. One can evaluate the reliability of the positions of the peak frequencies by investigating the noise dependency, as explained in another study \cite{PhysRevE.95.061302}.

\section{Summary}\label{Sec:Summary}
In this paper, we introduced the analytic continuation tool \verb|SpM|. 
In \verb|SpM|, the basis that is insensitive to noise is automatically selected using the sparse modeling method. As a result, stable analytic continuation against noise is realized. 

The present version of \verb|SpM| only provides analytic continuation using the sparse modeling method. A comparison of results obtained with other methods, such as \Pade approximation and maximum entropy methods, would be useful for evaluating the reliability of the analytic continuation. In the near future, by implementing these functions, \verb|SpM| will be able to examine results from various viewpoints and be a more helpful tool for understanding dynamical physical properties at finite temperatures in many-body quantum systems. 

\section*{Acknowledgments}
We thank T. Kato for useful comments. This work was supported by JSPS KAKENHI grants No. 16K17735, No. 18H04301 (J-Physics), and No. 18H01158.
KY and YM were supported by Building of Consortia for the Development of Human Resources in Science and Technology, MEXT, Japan. 
\section*{References}
\bibliographystyle{elsarticle-num}
\bibliography{spm.bib}

\begin{thebibliography}{10}
\expandafter\ifx\csname url\endcsname\relax
  \def\url#1{\texttt{#1}}\fi
\expandafter\ifx\csname urlprefix\endcsname\relax\def\urlprefix{URL }\fi
\expandafter\ifx\csname href\endcsname\relax
  \def\href#1#2{#2} \def\path#1{#1}\fi

\bibitem{Mahan}
G.~D. Mahan, Many-Particle Physics, Springer US, 2000.

\bibitem{gubernatis_kawashima_werner_2016}
J.~Gubernatis, N.~Kawashima, P.~Werner, Quantum Monte Carlo Methods: Algorithms
  for Lattice Models, Cambridge University Press, 2016.
\newblock \href {http://dx.doi.org/10.1017/CBO9780511902581}
  {\path{doi:10.1017/CBO9780511902581}}.

\bibitem{PhysRevB.61.5147}
K.~S.~D. Beach, R.~J. Gooding, F.~Marsiglio,
  \href{https://link.aps.org/doi/10.1103/PhysRevB.61.5147}{Reliable pad\'e
  analytical continuation method based on a high-accuracy symbolic computation
  algorithm}, Phys. Rev. B 61 (2000) 5147--5157.
\newblock \href {http://dx.doi.org/10.1103/PhysRevB.61.5147}
  {\path{doi:10.1103/PhysRevB.61.5147}}.
\newline\urlprefix\url{https://link.aps.org/doi/10.1103/PhysRevB.61.5147}

\bibitem{PhysRevB.93.075104}
J.~Sch\"ott, I.~L.~M. Locht, E.~Lundin, O.~Gr\aa{}n\"as, O.~Eriksson,
  I.~Di~Marco,
  \href{https://link.aps.org/doi/10.1103/PhysRevB.93.075104}{Analytic
  continuation by averaging pad\'e approximants}, Phys. Rev. B 93 (2016)
  075104.
\newblock \href {http://dx.doi.org/10.1103/PhysRevB.93.075104}
  {\path{doi:10.1103/PhysRevB.93.075104}}.
\newline\urlprefix\url{https://link.aps.org/doi/10.1103/PhysRevB.93.075104}

\bibitem{MaxEnt1990}
R.~Bryan, \href{https://doi.org/10.1007/BF02427376}{Maximum entropy image
  reconstruction - general algorithm}, Eur Biophys J 18 (1990) 165.
\newline\urlprefix\url{https://doi.org/10.1007/BF02427376}

\bibitem{JARRELL1996133}
M.~Jarrell, J.~Gubernatis,
  \href{http://www.sciencedirect.com/science/article/pii/0370157395000747}{Bayesian
  inference and the analytic continuation of imaginary-time quantum monte carlo
  data}, Physics Reports 269~(3) (1996) 133 -- 195.
\newblock \href
  {http://dx.doi.org/https://doi.org/10.1016/0370-1573(95)00074-7}
  {\path{doi:https://doi.org/10.1016/0370-1573(95)00074-7}}.
\newline\urlprefix\url{http://www.sciencedirect.com/science/article/pii/0370157395000747}

\bibitem{PhysRevB.81.155107}
O.~Gunnarsson, M.~W. Haverkort, G.~Sangiovanni,
  \href{https://link.aps.org/doi/10.1103/PhysRevB.81.155107}{Analytical
  continuation of imaginary axis data using maximum entropy}, Phys. Rev. B 81
  (2010) 155107.
\newblock \href {http://dx.doi.org/10.1103/PhysRevB.81.155107}
  {\path{doi:10.1103/PhysRevB.81.155107}}.
\newline\urlprefix\url{https://link.aps.org/doi/10.1103/PhysRevB.81.155107}

\bibitem{PhysRevB.57.10287}
A.~W. Sandvik,
  \href{https://link.aps.org/doi/10.1103/PhysRevB.57.10287}{Stochastic method
  for analytic continuation of quantum monte carlo data}, Phys. Rev. B 57
  (1998) 10287--10290.
\newblock \href {http://dx.doi.org/10.1103/PhysRevB.57.10287}
  {\path{doi:10.1103/PhysRevB.57.10287}}.
\newline\urlprefix\url{https://link.aps.org/doi/10.1103/PhysRevB.57.10287}

\bibitem{PhysRevE.81.056701}
S.~Fuchs, T.~Pruschke, M.~Jarrell,
  \href{https://link.aps.org/doi/10.1103/PhysRevE.81.056701}{Analytic
  continuation of quantum monte carlo data by stochastic analytical inference},
  Phys. Rev. E 81 (2010) 056701.
\newblock \href {http://dx.doi.org/10.1103/PhysRevE.81.056701}
  {\path{doi:10.1103/PhysRevE.81.056701}}.
\newline\urlprefix\url{https://link.aps.org/doi/10.1103/PhysRevE.81.056701}

\bibitem{PhysRevE.94.063308}
A.~W. Sandvik,
  \href{https://link.aps.org/doi/10.1103/PhysRevE.94.063308}{Constrained
  sampling method for analytic continuation}, Phys. Rev. E 94 (2016) 063308.
\newblock \href {http://dx.doi.org/10.1103/PhysRevE.94.063308}
  {\path{doi:10.1103/PhysRevE.94.063308}}.
\newline\urlprefix\url{https://link.aps.org/doi/10.1103/PhysRevE.94.063308}

\bibitem{Yoon2018}
H.~Yoon, J.-H. Sim, M.~J. Han,
  \href{https://link.aps.org/doi/10.1103/PhysRevB.98.245101}{Analytic
  continuation via domain knowledge free machine learning}, Phys. Rev. B 98
  (2018) 245101.
\newblock \href {http://dx.doi.org/10.1103/PhysRevB.98.245101}
  {\path{doi:10.1103/PhysRevB.98.245101}}.
\newline\urlprefix\url{https://link.aps.org/doi/10.1103/PhysRevB.98.245101}

\bibitem{PhysRevE.95.061302}
J.~Otsuki, M.~Ohzeki, H.~Shinaoka, K.~Yoshimi,
  \href{https://link.aps.org/doi/10.1103/PhysRevE.95.061302}{Sparse modeling
  approach to analytical continuation of imaginary-time quantum monte carlo
  data}, Phys. Rev. E 95 (2017) 061302.
\newblock \href {http://dx.doi.org/10.1103/PhysRevE.95.061302}
  {\path{doi:10.1103/PhysRevE.95.061302}}.
\newline\urlprefix\url{https://link.aps.org/doi/10.1103/PhysRevE.95.061302}

\bibitem{SpM}
URL https://github.com/SpM-lab/SpM.

\bibitem{Boyd:2011:DOS:2185815.2185816}
S.~Boyd, N.~Parikh, E.~Chu, B.~Peleato, J.~Eckstein,
  \href{http://dx.doi.org/10.1561/2200000016}{Distributed optimization and
  statistical learning via the alternating direction method of multipliers},
  Found. Trends Mach. Learn. 3~(1) (2011) 1--122.
\newblock \href {http://dx.doi.org/10.1561/2200000016}
  {\path{doi:10.1561/2200000016}}.
\newline\urlprefix\url{http://dx.doi.org/10.1561/2200000016}

\bibitem{10.2307/2346178}
R.~Tibshirani, \href{http://www.jstor.org/stable/2346178}{Regression shrinkage
  and selection via the lasso}, Journal of the Royal Statistical Society.
  Series B (Methodological) 58~(1) (1996) 267--288.
\newline\urlprefix\url{http://www.jstor.org/stable/2346178}

\bibitem{PhysRevB.92.060509}
A.~Reymbaut, D.~Bergeron, A.-M.~S. Tremblay,
  \href{https://link.aps.org/doi/10.1103/PhysRevB.92.060509}{Maximum entropy
  analytic continuation for spectral functions with nonpositive spectral
  weight}, Phys. Rev. B 92 (2015) 060509.
\newblock \href {http://dx.doi.org/10.1103/PhysRevB.92.060509}
  {\path{doi:10.1103/PhysRevB.92.060509}}.
\newline\urlprefix\url{https://link.aps.org/doi/10.1103/PhysRevB.92.060509}

\bibitem{PhysRevB.98.205102}
J.-H. Sim, M.~J. Han,
  \href{https://link.aps.org/doi/10.1103/PhysRevB.98.205102}{Maximum quantum
  entropy method}, Phys. Rev. B 98 (2018) 205102.
\newblock \href {http://dx.doi.org/10.1103/PhysRevB.98.205102}
  {\path{doi:10.1103/PhysRevB.98.205102}}.
\newline\urlprefix\url{https://link.aps.org/doi/10.1103/PhysRevB.98.205102}

\bibitem{PhysRevLett.112.070603}
A.~Decelle, F.~Ricci-Tersenghi,
  \href{https://link.aps.org/doi/10.1103/PhysRevLett.112.070603}{Pseudolikelihood
  decimation algorithm improving the inference of the interaction network in a
  general class of ising models}, Phys. Rev. Lett. 112 (2014) 070603.
\newblock \href {http://dx.doi.org/10.1103/PhysRevLett.112.070603}
  {\path{doi:10.1103/PhysRevLett.112.070603}}.
\newline\urlprefix\url{https://link.aps.org/doi/10.1103/PhysRevLett.112.070603}

\bibitem{doi:10.7566/JPSJ.84.024801}
S.~Yamanaka, M.~Ohzeki, A.~Decelle,
  \href{https://doi.org/10.7566/JPSJ.84.024801}{Detection of cheating by
  decimation algorithm}, Journal of the Physical Society of Japan 84~(2) (2015)
  024801.
\newblock \href {http://arxiv.org/abs/https://doi.org/10.7566/JPSJ.84.024801}
  {\path{arXiv:https://doi.org/10.7566/JPSJ.84.024801}}, \href
  {http://dx.doi.org/10.7566/JPSJ.84.024801}
  {\path{doi:10.7566/JPSJ.84.024801}}.
\newline\urlprefix\url{https://doi.org/10.7566/JPSJ.84.024801}

\end{thebibliography}

\end{document}